\documentclass[aps,prd,showpacs]{revtex4}
\usepackage{graphics,graphicx}
\usepackage{amsfonts}
\usepackage{amssymb}

\newcommand{\sect}[1]{Sec.\,#1}

\def\nhh{\hspace*{-0.3em}}
\def\cm{\hspace*{1cm}}
\def\inch{\hspace*{1in}}

\def\Jl#1#2{{\it #1\/} {\bf #2},\ }

\def\ApJ#1 {\Jl{Astroph. J.}{#1}}
\def\CQG#1 {\Jl{Class. Quantum Grav.}{#1}}
\def\DAN#1 {\Jl{Dokl. AN SSSR}{#1}}
\def\GC#1 {\Jl{Grav. \& Cosmol.}{#1}}
\def\GRG#1 {\Jl{Gen. Rel. Grav.}{#1}}
\def\JETF#1 {\Jl{Zh. Eksp. Teor. Fiz.}{#1}}
\def\JETP#1 {\Jl{Sov. Phys. JETP}{#1}}
\def\JHEP#1 {\Jl{JHEP}{#1}}
\def\JMP#1 {\Jl{J. Math. Phys.}{#1}}
\def\NPB#1 {\Jl{Nucl. Phys.}{B\ #1}}
\def\NP#1 {\Jl{Nucl. Phys.}{#1}}
\def\PLA#1 {\Jl{Phys. Lett.}{#1A}}
\def\PLB#1 {\Jl{Phys. Lett.}{#1B}}
\def\PRD#1 {\Jl{Phys. Rev.}{D\ #1}}
\def\PRL#1 {\Jl{Phys. Rev. Lett.}{#1}}

\def\lal{&& {}\nhh}
\def\eq{Eq.\,}

\def\beq{\begin{equation}}
\def\eeq{\end{equation}}
\def\bear{\begin{eqnarray}}
\def\bearr{\begin{eqnarray} \lal}
\def\ear{\end{eqnarray}}
\def\earn{\nonumber \end{eqnarray}}
\def\nn{\nonumber\\ {}}

\def\nnn{\nonumber\\ \lal }

\def\yy{\\[5pt] {}}

\def\eql{&\! = &\!}

\def\dst{\displaystyle}

\def\fracd#1#2{{\dst\frac{#1}{#2}}}

\def\Half{{\fracd{1}{2}}}

\def\e{{\,\rm e}}
\def\d{\partial}

\def\sign{\mathop{\rm sign}\nolimits}

\def\const{{\rm const}}
\def\eps{\varepsilon}

\def\DAL{\mathop{\raisebox{-.5pt}{$\Box$}}\nolimits}

\def\mn{_{\mu\nu}}
\def\MN{^{\mu\nu}}

\def\cK{{\cal K}}
\def\cV{{\cal V}}

\def\kappa{\varkappa}

\def\wt{\widetilde}
\def\tg{{\wt g}}
\def\tR{{\wt R}}

\def\M{{\mathbb M}}

\def\Lambdaef{\Lambda_{\rm eff}}

\def\sss{\scriptscriptstyle}

\def\mD{m_{\sss D}}

\begin{document}

\title{Diversity of universes created by pure gravity}

\author{K.A. Bronnikov}
\affiliation
    {Center for Gravitation and Fundamental Metrology, VNIIMS, 46 Ozyornaya St., Moscow, Russia;
     Institute of Gravitation and Cosmology, PFUR, 6 Miklukho-Maklaya St., Moscow 117198,
     Russia.//
         E-mail: kb20@yandex.ru}

\author{R.V. Konoplich}
\affiliation
    {Department of Physics, New York University,
New York, NY 10003, Physics Department, USA; Manhattan College,
Riverdale, New York, NY 10471, USA,//
 E-mail: rk60@nyu.edu}

\author{S.G. Rubin}
\affiliation
    {Moscow Engineering Physics Institute, 31 Kashirskoe Sh., Moscow 115409,
    Russia.//
    E-mail: sergeirubin@list.ru}

\begin{abstract}
 {We show that a number of problems of modern cosmology may be solved in the
  framework of multidimensional gravity with high-order curvature
  invariants, without invoking other fields. We use a method employing a
  slow-change approximation, able to work with rather a general form of the
  gravitational action, and consider Kaluza-Klein type space-times with one
  or several extra factor spaces. A vast choice of effective theories
  suggested by the present framework may be stressed:  even if the initial
  Lagrangian is entirely fixed, one obtains quite different models for
  different numbers, dimensions and topologies of the extra factor spaces.
  As examples of problems addressed we consider (i) explanation of the
  present accelerated expansion of the Universe, with a reasonably small
  cosmological constant, and the problem of its fine tuning is considered
  from a new point of view; (ii) the mechanism of closed wall production in
  the early Universe; such walls are necessary for massive primordial black
  hole formation which is an important stage in some scenarios of cosmic
  structure formation; (iii) sufficient particle production rate at the end
  of inflation; (iv) it is shown that our Universe may
  contain spatial domains with a macroscopic size of extra dimensions.
  We also discuss chaotic attractors appearing at possible
  nodes of the kinetic term of the effective scalar field Lagrangian.  }

\pacs{04.50.+h; 98.80.-k; 98.80.Cq}

\end{abstract}
\maketitle

\section{Introduction}

  Owing to the advances of modern cosmology, our Universe is now treated as
  a dynamically developing object. Its early period is rather adequately
  described by scenarios containing an inflationary stage. Its future is
  yet rather uncertain and depends both on the parameters of the underlying
  theory and on the initial data. Some other problems of principle are
  also yet to be solved. Among them one can mention the essence of dark
  energy, early formation of supermassive black holes (which is a necessary
  stage in some scenarios of cosmic structure formation), and sufficient
  particle production at the end of inflation.

  In the authors' view, multidimensional gravity may be taken as a basis for
  solving these and maybe other problems using a minimal set of postulates.
  The purpose of this paper is to demonstrate that multidimensional gravity
  alone leads to a diversity of low-energy theories owing to arbitrariness
  of the ``input'' parameters in the effective action and to possible
  different initial conditions concerning both the metric and the topology
  of extra dimensions. Both of them are thought to be produced by quantum
  fluctuations at high energies. Different effective theories take place
  even with fixed parameters of the original Lagrangian.

  Thus our starting point is that {\sl the total space-time dimension $D$ is
  greater than four. The %particular value of $D$ and the
  topological properties of space-time are determined by quantum fluctuations and may
  vary, leading to drastically different universes.}

  Our standpoint is close to that of chaotic inflation, according to
  which infinitely many universes are permanently created with a scalar
  (inflaton) field corresponding to different values of its potential. If
  the inflaton potential is simple, the universes are similar to each other.
  A more complex form of the potential gives rise to universes with
  different properties, and the situation begins to resemble the prediction
  of string theory known as the landscape concept: the total number of
  different vacua in heterotic string theory is about $10^{1500}$
  \cite{Landscape0}); more realistic, de Sitter vacua are considered in
  \cite{Landscape1}. The number of possible different universes is then
  huge but finite. Moreover, the concept of a random potential \cite{Random}
  leads to an infinite number of universes with various properties. We make
  a step further and try to ascribe the origin of such potentials to
  multidimensional gravity.

  The idea of extra dimensions, tracing back to the pioneering works of T.
  Kaluza and O. Klein, has now become a common ingredient of practically all
  attempts to unify all physical interactions. We here do not rely on a
  particular unification theory but are rather trying to explore
  consequences of the very idea of extra dimensions.

  Another crucial point is that {\sl the pure gravitational action contains
  curvature-nonlinear terms}. It is a direct consequence of quantum field
  theory in curved space-time \cite{GMM,BD} and should not be considered as
  an independent postulate.

  Here we show that our approach, without need for fields
  other than gravity, are able to produce such different structures as
  inflationary (or simply accelerating) universes, brane worlds, black holes
  etc. The role of scalar fields is played by the metric components of extra
  dimensions.

  In the present paper, we will be concerned with cosmological aspects of
  this approach. In our previous paper \cite{BR06} it has been shown that
  many models of interest, able to describe an accelerated expansion of the
  Universe, may be obtained in this approach even with the simplest
  Lagrangians, quadratic in the multidimensional curvature. Here we widen
  our scope and, in addition to modern acceleration, consider some problems
  of the early Universe. We show, in particular, that the conditions
  required for the onset and development of inflation may also emerge in the
  low-energy limit of the original multidimensional theory.

  Inflation is, in general, a very attractive phenomenological idea,
  supported by observations, but its foundation on a fundamental level
  remains uncertain. The inflationary Universe paradigm appeared in the
  early 80s and suggested solutions to a number of long-standing problems of
  relativistic cosmology \cite{infl-80}, and a great number of various
  inflationary scenarios have been created since then (for reviews see e.g.
  \cite{infl+, Lyth}). Modern observations \cite{Lyth} confirm the most
  valuable predictions of inflation with increasing accuracy but strongly
  constrain specific inflationary models. For example, according to the
  observations, the spectral index of curvature perturbations could be less
  than unity. It makes doubtful the Hybrid model of inflation \cite{Hyb} in
  its simplest realization, which had seemed to be promising due to the
  absence of small parameters. On the other hand, an inflaton field with the
  simplest, quadratic potential can conform to the observations, but the
  smallness of its parameters still needs an explanation.

  A common feature of almost all inflationary models is the existence of a
  scalar field, called the inflaton, or a set of such fields, possessing a
  nontrivial potential. There are a number of ways of explaining the origin
  of such fields: supergravity, string and brane ideas, nonlinear gravity,
  extra dimensions etc. So the nature of a true inflaton field (or fields)
  is yet to be understood. We wish to argue that multidimensional gravity
  creates such fields most naturally, with a minimum number of postulates.

  Other challenging problems are those related to fine tuning, required to
  explain the actual properties of our Universe. Thus, small parameters are
  necessary to provide the smallness of temperature fluctuations of the CMB;
  an extremely small parameter is required to explain the observed value of
  dark energy density. We here show that such fine-tuning problems may be
  reduced to the problem of the number of extra dimensions.

  The paper is organized as follows.
  In \sect II we briefly reproduce our basic approach developed in
  \cite{BR06}, in the simplest Kaluza-Klein framework with one extra factor
  space of arbitrary dimension. Following \cite{BR06}, it is shown that
  nonlinear multidimensional gravity in its low-energy limit is equivalent
  to a certain scalar-tensor theory. In \sect III we discuss two problems in
  this framework: one is related to multiple domain wall formation at the
  inflationary stage of the Universe evolution, which can explain the
  existence of primordial supermassive black holes; the other problem
  concerns intensive particle and entropy production in the post-inflationary
  period. \sect IV is devoted to a discussion of a feature of interest of
  the framework under study, that the kinetic term of the effective scalar
  field(s) contains, as a factor, a function of the field itself, which can
  possess zeros; we study the possible state of the system near such a zero.
  \sect 5 discusses the dependence of the effective low-energy theory on the
  structure of compact extra dimensions. In particular, the number of
  effective scalar fields coincides with the number of extra factor spaces,
  therefore the effective theories may be quite diverse even with fixed
  values of the parameters of the initial Lagrangian.

  We would like to stress that our purpose here is to demonstrate the power
  of the approach rather than create particular models to be confronted
  with observations; the latter will be a subject of future work.

\section{Basic equations, $F(R)$ gravity}

    Let the space-time have the structure $\M_D = \M_d \times \M_4$, where
    the extra factor space $\M_d$ is of arbitrary dimension $d$ and
    assumed to be a space of positive or negative constant curvature $k =
    \pm 1$. Consider the action
\beq                                            \label{S1}
    S = \Half \mD^{D-2} \int \sqrt{^{D}g}\, d^{D}x\, [F(R) + L_m]
\eeq
    and the $D$-dimensional metric
\beq                                                            \label{ds_D}
     ds^{2} = g_{\mu\nu}(x) dx^{\mu}dx^{\nu}
                        + \e^{2\beta(x)}h_{ab}dx^{a}dx^{b}
\eeq
    where $(x)$ means the dependence on $x^\mu$, the coordinates of $\M_4$;
    $h_{ab}$ is the $x$-independent metric in $\M_d$. The choice
    (\ref{ds_D}) is used in many studies, e.g., \cite{Zhuk,Holman,Majumdar}.
    It is quite general in the low-energy limit used throughout this paper.
    Indeed, suppose we start with a more general ansatz,
\beq\label{111}
    ds^{2} = g_{\mu\nu}(x) dx^{\mu}dx^{\nu} +
                 \e^{2\beta(x,y)}h_{ab}dy^{a}dy^{b}
\eeq
    and that the extra factor space is compact.
%%(we will below comment on when it is possible and thus confirm the
%% validity of our approach).
    Then we can use a Fourier expansion in the corresponding orthonormalized
    set of functions $Y_p(y)$ (e.g., $Y_p$ are $d$-dimensional spherical
    functions if $\M_d$ has the topology of a sphere):
\[
    \e^{2\beta (x,y)}  = \sum_p Y_p(y) B_p (x).
\]
    The mode with the lowest energy corresponds to $Y_0 = \const$, others
    are designated as excited Kaluza-Klein modes, and each of them can be
    seen as a four-dimensional scalar field satisfying the four-dimensional
    Klein-Gordon equation with a nonzero squared mass. They represent the
    so-called KK tower and have high excitation energies inversely
    proportional to a characteristic size of the extra dimensions,
    see, e.g., \cite{Majumdar99}.
    We do not consider them because we are working in the low-energy
    limit. It means that we may substitute
\[
    \e^{2\beta (x,y)}  \to B_0(x) = \e^{2\beta(x)}
\]
    from the very beginning.

    In (\ref{S1}), the matter Lagrangian $L_m$ is included for generality
    and is not used in the following sections. Capital Latin indices cover
    all $D$ coordinates, small Greek ones cover the coordinates of $\M_4$
    and $a, b, \ldots$ the coordinates of $\M_d$. The $D$-dimensional Planck
    mass $\mD$ does not necessarily coincide with the conventional Planck
    scale $m_4$; $\mD$ is, to a certain extent, an arbitrary parameter, but
    on observational grounds it must not be smaller than a few TeV.

    The Ricci scalar can be written in the form
\bear                                   \label{R-decomp}
        R \eql R_{4} + \phi + f_{\rm der},
\nn
        \phi \eql kd(d-1) \mD^2 \e^{-2\beta(x)}
\nn
      f_{\rm der} \eql 2d g^{\mu\nu}\nabla_{\mu}\nabla_{\nu}\beta
                + d(d+1) (\d\beta)^2,
\ear
    where $(\d\beta)^2 = g\MN \d_\mu \beta \d_\mu \beta$.
    The {\sl slow-change approximation\/}, employed in \cite{BR06}, assumes
    that all quantities are slowly varying, i.e., it considers each derivative
    $\d_{\mu}$ (including those in the definition of $R_4$) as an expression
    containing a small parameter $\eps$, so that
\beq
        |\phi| \gg |R_4|,\ |f_{\rm der}|.
\eeq
    As shown in \cite{BR06}, this approximation even holds in any
    inflationary model whose characteristic energy scale is far below the
    Planck scale $\mD$ (to say nothing of the modern epoch).
    Thus, the GUT scale, which is common in inflationary models, is $m_{\rm
    GUT} \sim 10^{-3} m_4$, which means that primordial inflation may
    be well described in the present framework if $\mD \sim m_4$.

    In this approximation, using a Taylor decomposition for $F(R) = F(\phi +
    R_{4} + f_{\rm der})$ and integrating out the extra dimensions, we
    obtain up to $O(\eps^2)$
\bearr
     S = \Half \cV_d\, \mD^2                                 \label{act3}
          \int \sqrt{^{4}g}\,d^{4}x\, \e^{d \beta}
             [F'(\phi)R_4 + F(\phi) + F'(\phi) f_{\rm der} + L_m],
\ear
    where $^{4}g = |\det(g\mn)|$ and $\cV_d$ is the volume of a compact
    $d$-dimensional space of unit curvature. The expression (\ref{act3}) is
    typical of a scalar-tensor theory (STT) of gravity in a Jordan frame.

    To find stationary points, it is helpful to pass on to the Einstein
    frame. After the conformal mapping
\beq                                                       \label{trans-g}
    g\mn \ \mapsto \tg\mn = |f(\phi)| g\mn, \cm
            f(\phi) =  \e^{d\beta}F'(\phi),
\eeq
    (the tilde marks quantities obtained from or with $\tg\mn$),
    the action (\ref{act3}) may be brought to the form
\bear
     S \eql \frac{\cV[d]}{2} \mD^2 \int d^{4}x\, \sqrt{\tg}\, (\sign F') L,
\nn
     L \eql \tR_4 + \Half K_{\rm Ein}(\phi) (\d\phi)^2
                        - V_{\rm Ein}(\phi) + {\wt L}_m,   \label{Lgen}
\cm
          {\wt L}_m = (\sign F')\frac{\e^{-d\beta}}{F'(\phi)^2} L_m;
\\
     K_{\rm Ein}(\phi) \eql                                \label{KE}
        \frac{1}{2\phi^2} \left[
            6\phi^2 \biggl(\frac{F''}{F'}\biggr)^2\!
            -2 d \phi \frac{F''}{F'} + \Half d (d + 2)\right],
\\
     V_{\rm Ein}(\phi) \eql - (\sign F')
        \left[\frac{|\phi|\mD^{-2}}{d (d -1)}\right]^{d/2}
                \frac{F(\phi)}{F'(\phi)^2 }                \label{VE}
\ear
    In (\ref{Lgen})--(\ref{VE}), the indices are raised and lowered
    with $\tg\mn$; everywhere $F = F(\phi)$ and $F' = dF/d\phi$.
    All quantities of orders higher than $O(\eps^2)$ are neglected.

    The quantity $\phi$ may be interpreted as a scalar field with the
    dimensionality $\mD^2$, see (\ref{R-decomp}). In what follows (if not
    indicated otherwise) we put $\mD =1$.

    In fact, the function $F(\phi)$ represents an infinite power series
    inevitably caused by quantum corrections. Most frequently, only a few
    first terms are considered, and the simplest among them is quadratic,
\beq                                                     \label{quadr}
            F(R) = R + cR^{2} - 2\Lambda.
\eeq
    It has been shown \cite{BR06,Zhuk} that, in the case of a negative
    curvature of the extra dimensions, the potential (\ref{VE}) for
    $F(R)$ of the form (\ref{quadr}) may possess a minimum. Here we will
    also use the more general form
\beq                            \label{quartic}
        F(R) = R + cR^{2} + w_{1}R^{3} + w_{2}R^{4} - 2\Lambda
\eeq
    and show that it leads to new nontrivial results.

\section{Two applications to cosmological problems}

    Multidimensional nonlinear gravity presents a way to create universes
    with different sets of properties, and many of them are of considerable
    interest. Thus, some models with a small size of extra dimensions are
    metastable, and the vacuum decay probability depends on the energy
    difference between two minima of the potential
    \cite{Coleman,Konoplich,Kolb90,Lav,Dunne}. When, however, we seek a model
    of the modern Universe with an extremely small energy density,
    $\rho \sim 10^{-123} m_4^4$, we may not bother about vacuum percolation.

    Below in this section we discuss two particular problems existing in the
    description of the initial and final stages of primordial inflation
    and show how they can be addressed in the presently discussed framework.

    One of the problems is connected with multiple domain wall production if
    the Universe nucleation occurs not too far from a maximum or a saddle
    point of the inflaton potential. Another problem concerns sufficient
    matter creation from inflaton oscillations at the end of inflation.

    For certainty, we will everywhere treat the Einstein conformal frame, in
    which the Lagrangian has the form (\ref{Lgen}), as the physical frame
    used to interpret the observations. It should be stressed that it is
    only one of possible interpretations, see a discussion in \cite{BR06}.
    We also put $\mD = m_4$ in estimates.

\subsection{Multiple closed wall and black hole production}

    Some inflationary models suppose creation of our Universe either near a
    maximum of the potential or near its saddle point(s) to obtain slow
    rolling providing a sufficient number of e-foldings, see, e.g.,
    \cite{Dvali, Racetrack}. Our assumptions lead to effective potential
    (\ref{VE}) and also make it possible to obtain different potentials possessing
    maxima at a high enough level and minima able to describe the end of
    inflation.
    %One of them is represented in Fig.\,\ref{potkinMIN}.

    There is an important problem, usually missed in discussions and
    connected with the effect of quantum fluctuations of the scalar field
    during inflation. Let us discuss it in this section. The evolving
    inflaton field may be split into classical parts, governed by the
    classical equation of motion, and quantum fluctuations \cite{Star}. To
    consider the latter, let us approximate the potential near its maximum
    as
\beq
        V = V_0 - \frac{m^2}{2} \phi ^{2},      \label{Vapprox}
\eeq
    where the maximum is assumed, without loss of generality, at $\phi
    =0$.
    Then the density of probability to find a certain field value $\phi$ has
    the form \cite{Book1} (adapted to classical motion near a
    maximum rather than a minimum)
\beq
        dP(\phi, T;\phi_{\rm in}, 0) = d\phi \sqrt{\frac{a}{\pi
            ( \e^{2\mu T}-1) }} \exp \left[ -a\frac{( \phi
              - \phi_{\rm in}\e^{\mu T})^2 }{\e^{2\mu T}-1} \right].
\eeq
    Here $a = \mu /\sigma ^{2}$, $\mu \equiv m^{2}/(3H)$ and $\sigma
    = H^{3/2}/(2\pi)$ where the Hubble parameter is
\[
        H\simeq \sqrt{8\pi V_0/(3m_{4})}.
\]
    Let the initial field value be $\phi_{\rm in} > 0$. Then the average
    field value increases with time, ultimately reaching a minimum of the
    potential at some $\phi =\phi _{+}>0$. The greater part of space will be
    filled with $\phi$ at this value. Meanwhile, some quantum fluctuations
    could jump over the maximum, and the average field value representing
    this fluctuation tends to another minimum of the potential, $\phi_{-}<0$.
    Thus some spatial domains are characterized by $\phi_{-} < 0$. There
    inevitably forms a wall between such domains and the ``outer'' space
    with $\phi = \phi_+$ \cite{Book1, PBH}.

\begin{figure}                     %% fig 1
    \includegraphics[width=0.5\textwidth]{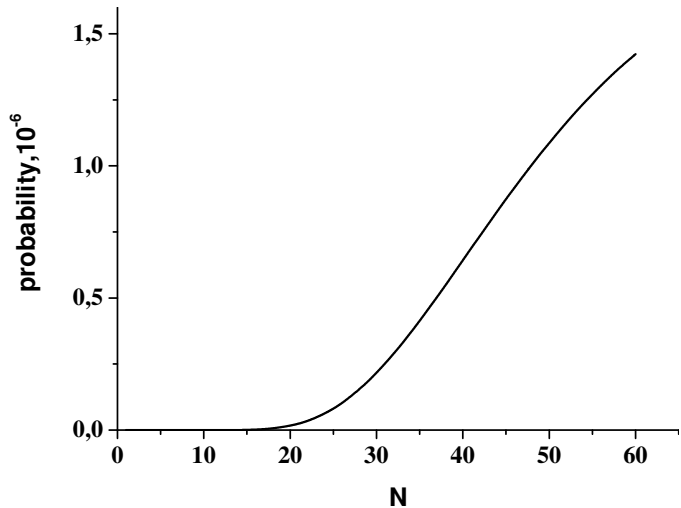}\\
\caption
  {\small Part of space occupied by another vacuum state, depending of time
  measured by e-folds. The initial field value is
  $\phi_{\rm in} = \phi_{\max} + 3H$, where  $\phi^{\max}$ is the field
  value at a maximum of the potential. The ``inflaton mass'' is chosen to be
  $m=0.3H$, where $H$ is the Hubble parameter at the top of the
  potential.}
\label{ratio}
\end{figure}

    Thus ``dangerous'' values of fluctuations are those for which
    $\phi \leq 0$, and it is instructive to calculate the probability of
    their nucleation. The latter strictly depends on the field value $\phi
    _{\rm in}$ at the moment of creation of our Universe.
    Fig.\,\ref{ratio} represents the probability
\beq
       P_{0}(\phi _{\rm in},T)=\int_{\phi = -\infty }^{\phi = 0}
                     dP (\phi, T; \phi_{\rm in}, 0)
\eeq
     to find the field value $\phi \leq 0$ at some point of space for
     reasonable values of the parameters. It determines a ratio of spatial
     volumes with different signs of the field.  This probability (which
     makes sense for specified values of the model parameters) is highly
     sensitive to the initial field value $\phi_{\rm in}$: the closer it is
     to the potential maximum, the greater part of the Universe will be
     covered with walls at the final stage.

    If the walls surround only a sufficiently small part of space, then
    supermassive black holes, which are formed from the walls \cite{Book1}
    could explain the early formation of quasars \cite{DER}. An increased
    fraction of space inside the domain walls leads to an increased number
    of such primordial black holes as well as their increased individual
    masses; the BH contribution to the dark matter density then becomes
    unacceptably large \cite{PBH}, or, in another scenario, there emerges a
    wall-dominated universe \cite{Zeld}. Therefore, the mechanism
    considered should be taken into account in the construction of
    particular inflationary scenarios like New Inflation \cite{Steinhardt}.

\subsection {Effective particle production after inflation}

    According to \cite{DolgovAbbott}, quick oscillations of the inflaton
    field immediately after the end of the inflationary stage are necessary
    for particle production which should lead to the observable amount of
    matter. It is known \cite{KLS, Shtanov} that if the inflaton coupling
    to matter fields is negligible, the mechanism of particle/entropy
    production and heating of the Universe could be ineffective. On the
    other hand, a strong coupling between the inflaton and matter fields,
    leading to large quantum corrections to the initial Lagrangian, would
    set to doubt the sufficiently small values of the input parameters
    needed for the very existence of the inflationary stage. The Hybrid
    inflationary model \cite{Hyb} successfully  solves this problem by
    incorporating one more, ``waterfall'', field. During inflation, the
    energy density slowly varies until the state reaches a bifurcation
    point. After that, the energy density quickly drops, producing quick
    oscillations near the bottom of the potential thus giving rise to the
    desirable multiple particle production.  Unfortunately, this promising
    model suffers overproduction of supermassive black holes discussed
    above, see \cite{PBH}.  Another possibility of effective particle
    creation (parametric resonance) was described in \cite{KLS}. Nonlinear
    multidimensional gravity yields one more mechanism.

\begin{figure}
\includegraphics[width=0.5\textwidth]{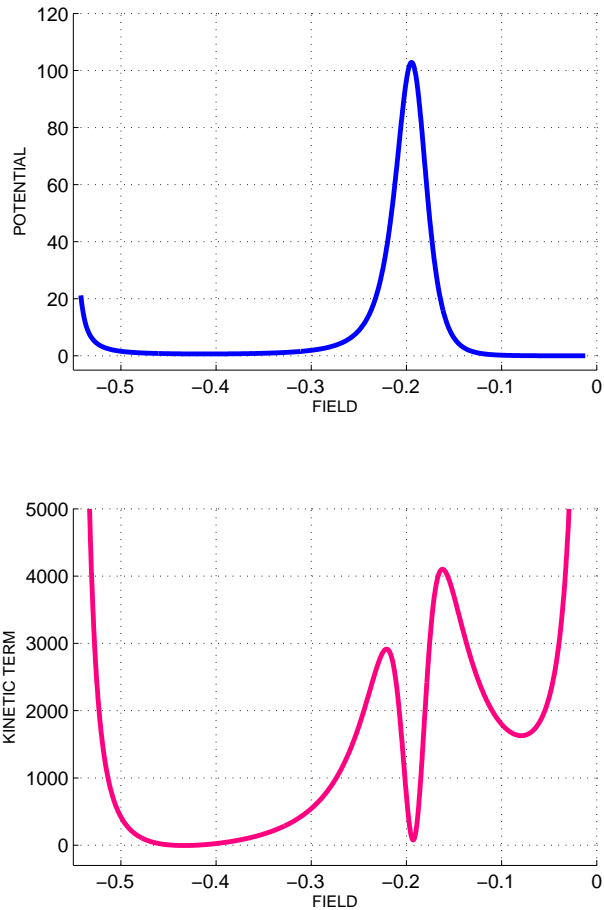}\\
\caption{The potential and kinetic terms in the effective 4D theory for the
initial theory (\ref{quartic}) with the parameters $d=3,\ c=6.0,\ w_1 =15,\
w_2 =12,\ \Lambda =2.$}\label{potkinMIN}
\end{figure}

    Consider the potential and kinetic terms for the effective Lagrangian
    (\ref{Lgen}) with the function $F(R)$ taken from (\ref{quartic}).
    Their plots for specific values of the parameters are represented in
    Fig.\,\ref{potkinMIN}. A nontrivial $\phi $ dependence of the kinetic
    term strongly affects the classical scalar field dynamics. The effect is
    especially strong when the minimum of the kinetic term approximately
    coincides with that of the potential, as is the case in
    Fig.\,\ref{potkinMIN}.

    To illustrate the situation, consider a toy model of a scalar
    field $\phi$ with
\bear                   \label{kinmodel}
    V(\phi ) \eql \frac12 m^2\phi^2
\nn
         K(\phi) \eql K_1\cdot (\phi - \phi_{\min})^2 + K_{\min},
           \cm  K_1,\ K_{\min} >0.
\ear
\begin{figure}                          %% Fig 3
\includegraphics[width=0.75\textwidth]{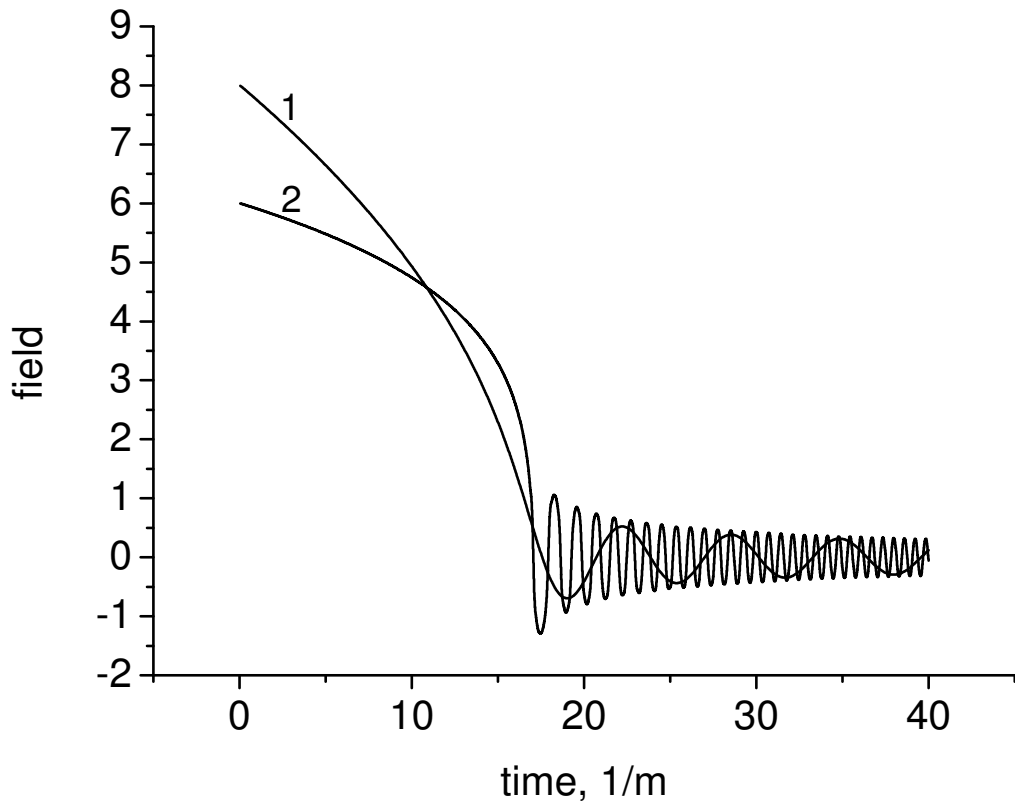}\\
\caption{Behaviour of the inflaton at the end of inflation for two
   models. 1. The standard kinetic term, $K=1/2$. 2. A kinetic term of the
   form (\ref{kinmodel}). The parameters $K_0 =0.1,\ \phi_{\min}=0.1,\
   K_{\min}=0.005$ are given in Planck units, and time is measured in the
   units of inverse mass of the inflaton field. Both plots display the final
   stage of inflation lasting about 60 e-folds.}\label{dynamics}
\end{figure}
 (The function $K(\phi)$ is positive and could be reduced to unity by the
 field redefinition $\phi\to \tilde\phi$, but this would be only another
 formulation of the same theory, less convenient for our purposes.)

    Numerical solutions to the corresponding classical equations
    for spatially flat FRW cosmologies
\bearr
    H^{2} =\frac{\kappa^{2}}{3}\left[
        \frac{1}{2}K(\phi)\dot{\phi}^{2} + V(\phi)\right]  ,
\nnn
    K(\phi)\left(\ddot{\phi} + 3H\dot{\phi}\right)
        + \frac{1}{2}K_{\phi}(\phi)\dot{\phi}^{2}
            + V_{\phi} (\phi) = 0
\ear
   (the index $\phi$ means $d/d\phi$, and $H$ is the Hubble parameter)
   are presented in Fig.\,\ref{dynamics} for quadratic potential (\ref{kinmodel})
   and two different kinetic terms: one with $K= 1/2$ and the other given by
   (\ref{kinmodel}), with parameter values given in the figure caption. Both
   plots display a few last e-folds of inflation and the consequent field
   oscillations.  The nontrivial kinetic term $K(\phi)$ equals 1/2 at the
   end of inflation (the first point of line crossing in
   Fig.\,\ref{dynamics}) for the chosen values of the parameters.
   Evidently, inflation in models with nontrivial kinetic terms starts at
   smaller energies as compared with the ordinary kinetic term, in agreement
   with the results obtained by Morris \cite{Morris} in the framework of
   inflation at an intermediate scale.

   An important observation is that in our case the frequency of inflaton
   oscillations is much larger than with the standard kinetic term, although
   inflation lasts for the same time. The resulting number of produced
   particles is proportional to this frequency, see the detailed review
   \cite{Dolgov} and an additional discussion in \cite{Book1}. This
   conclusion could be qualitatively supported by the following argument.

   When the inflation is over, the amplitude of inflaton oscillations is
   small on the Planck scale, and the effective kinetic term $K_{\rm
   eff}\sim K_{\min}$ is small due to the chosen values of the parameters.
   The effective Lagrangian, containing an interaction term of the inflaton
   and some other scalar field $\chi$
\beq        \label{Leff}
    L_{\rm eff}\simeq \frac12 K_{\min}\dot{\phi}^2 + g\phi\chi\chi,
\eeq
   can be transformed into the standard form by re-definition of the
   inflaton field:
\beq            \label{LeffStand}
    L_{\rm eff}\simeq\frac12 \dot{\phi}^2 +
            \frac{g}{\sqrt{K_{\min}}}\phi\chi\chi.
\eeq
   A small value of $K_{\min}$ increases the effective coupling constant
   (by an order of magnitude for the chosen values of the parameters) and
   surely leads to more intensive particle production. As a result, we
   overcome the difficulty discussed above: a slow motion during inflation
   may be reconciled with effective particle production right after
   inflation. Hence universes of this sort possess promising conditions for
   creation of complex structures.

\section{A non-classical state near $K =0$}

    In this section we discuss the possible state of a universe near a
    node of the kinetic term.
    Ref.\,\cite{Kroger} has indicated some consequences of singular points of
    kinetic terms in Einstein-scalar field theory. Recall that, in terms of
    \eq (\ref{Lgen}), $K(\phi) = K_{\rm Ein} > 0$ corresponds to a normal
    scalar field and $K(\phi) < 0$ to a phantom field, with negative kinetic
    energy. Nodes and poles of $K(\phi)$ are of special interest. Let us
    show that such singular points of the kinetic term may take place in
    nonlinear gravity with extra dimensions.

    To begin with, $F(R)$ theories considered in the previous sections
    produce only positive kinetic terms. Indeed, as is easily verified, the
    expression is square brackets in \eq (\ref{KE}) is always positive, its
    minimum value being $d(d+3)/3$.
\begin{figure}
    \includegraphics[width=0.5\textwidth]{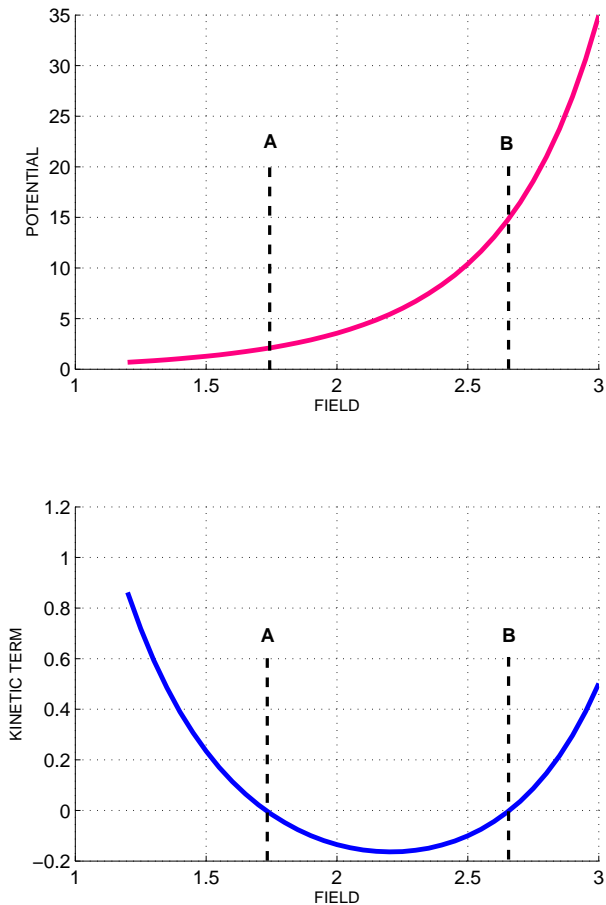}
\caption{The potential (\ref{VEKr}) and the kinetic term (\ref{KEKr}) for
     the model (\ref{SKr}) and $F(R)$ from (\ref{quadr}).
     The parameters are $d=3, c=-0.12, \Lambda =2,c_1 +c_2 =-5.5$, $c'=-0.626$.}\label{potkina}
\end{figure}
    However, if curvature-nonlinear terms in the Lagrangian exist due to
    quantum field effects, it seems reasonable to suppose that there can be
    contributions other than $F(R)$, and then the kinetic term of the
    effective theory is no more necessarily positive-definite. As an
    example, consider the simplest nonlinear theory (\ref{quadr})
    including additional nonlinear terms of quadratic order: the Ricci
    tensor squared $R_{AB}R^{AB}$ and the Kretschmann scalar $\cK =
    R_{ABCD}R^{ABCD}$, so that
\beq                                                  \label{SKr}
    S = \int d^{D}x\sqrt{^{D}g}\left[
            F(R) + c_{1} R_{AB}R^{AB} + c_{2}\cK \right].
\eeq

    Calculations for (\ref{SKr}) with $F(R)$ given by (\ref{quadr}) and the
    metric (\ref{ds_D}), similar to those made in \sect 2, lead to an
    effective Einstein-scalar theory with
\bear
     V_{\rm Ein}(\phi ) \eql -\sign(1+2c\phi) [ d(d-1) ]^{-d/2}
    \cdot | \phi|^{d/2}\frac{c'\phi^{2}+\phi -2\Lambda }{(1+2c\phi)^{2}},
\nnn  \cm                                   \label{VEKr}
    c' = c+\frac{c_{1}}{d}+\frac{2c_{2}}{d(d-1)},
\yy                                         \label{KEKr}
     K_{\rm Ein}(\phi ) \eql
          \frac{1}{\phi^{2}(1+2c\phi)^{2}}
            \left[ c^{2}\phi^{2} ( d^2 -2d +12) +d^{2}c\phi
    +\frac{1}{4}d(d+2)\right] + \frac{c_{1}+c_{2}}{2\phi (1+2c\phi)}.
\ear

   In Fig.\,\ref{potkina}, the dashed lines A and B mark nodes of the
   kinetic term in the case of quadratic gravity (\ref{quadr}). If a field
   fluctuation is produced to the left of line A, the field tends to the
   minimum of the potential located at $\phi=0$, corresponding to large
   extra dimensions. Some fluctuations are formed in the region between
   lines A and B, where the kinetic term is negative, the $\phi$ field is
   phantom and tends to climb the slope of the potential. If a fluctuation
   appears to the right of line B, it has a good opportunity to produce a
   universe with a nonzero cosmological constant. The ground state of such
   a universe is nontrivial and worth discussing in more detail.

   To this end, consider a toy model with the action
\beq
    S=\int d^{4}x\left[ \frac{1}{2}K(\phi )
        (\d \phi )^{2} - V(\phi )  \right]
\eeq
    with $K(\phi _{\rm crit})=0$ by definition and, without loss of
    generality, suppose that $\phi_{\rm crit}=0$. Then the generic behaviour
    of $K(\phi)$ and $V(\phi)$ at small $\phi$ is $K(\phi ) = k\phi +o(\phi
    )$ and $V(\phi ) = V(0) + h\phi + o(\phi )$, where we put $k > 0$, $h >
    0$ by analogy with point B in Fig.\,4. Thus, near the critical point,
    the field tends upward if it is to the left of it and downward if it is
    to the right of it. In other words, the critical point looks like an
    attractor. Nevertheless, there is no classical motion near the point
    $\phi =0$. Indeed, the classical equation for $\phi$ in curved space-time
    has the general form
\beq
    K(\phi )\DAL \phi  + \frac{1}{2}K_\phi (\d\phi )^{2}=-V_\phi,
    \label{main}
\eeq
    or, according to the above expressions for $K$ and $V$ at small $\phi$,
\beq
       k\phi \DAL\phi =-\left(h+\frac{1}{2}k(\d \phi )^{2}\right) +o(\phi).
    \label{Taylor}
\eeq
    Suppose that $(\d\phi)^2 >0$ (as is the case in cosmology when
    $\phi = \phi(t)$). Then the right-hand side of (\ref{Taylor}) is smaller
    than $-h < 0$ and does not tend to zero as $\phi \to =0$.
    Hence the second derivatives involved in $\DAL\phi$ tend to infinity as
    $\phi \to \phi_{\rm crit}=0$. One could improve the situation by
    invoking terms in the Lagrangian containing higher-order derivatives,
    making it possible to avoid infinite values of the derivatives, but it
    still remains unusual because a stationary state at the attractor is
    absent in any case. Indeed, if all derivatives are zero, the classical
    equation (\ref{Taylor}) with $h >0$ has no solution. It means that the
    kinetic energy of such a ground state cannot be zero. The problem
    obviously deserves a further study.

    One can notice that a classical solution to \eq (\ref{Taylor}) appears
    if $(\d\phi)^2 < 0$, i.e., the field gradient is spacelike. It may thus
    happen that near the critical point there inevitably appear spatial
    inhomogeneities, which may lead to an instability.

\section{Alternative number and structure of extra dimensions}

  The previous section discussed effective low-energy theories corresponding
  to the metric (\ref{ds_D}) for different choices of the initial action.
  It is not surprising that some values of the parameters are suitable for
  the description of a universe like ours. Additional opportunities appear
  when varying the structure of extra dimensions, which includes the number
  of extra factor spaces, their dimensionality and curvature.  Even if the
  initial Lagrangian is entirely specified, this additional freedom makes it
  possible to obtain low-energy effective Lagrangians drastically different
  from one another.

\begin{figure}
\includegraphics[width=0.5\textwidth]{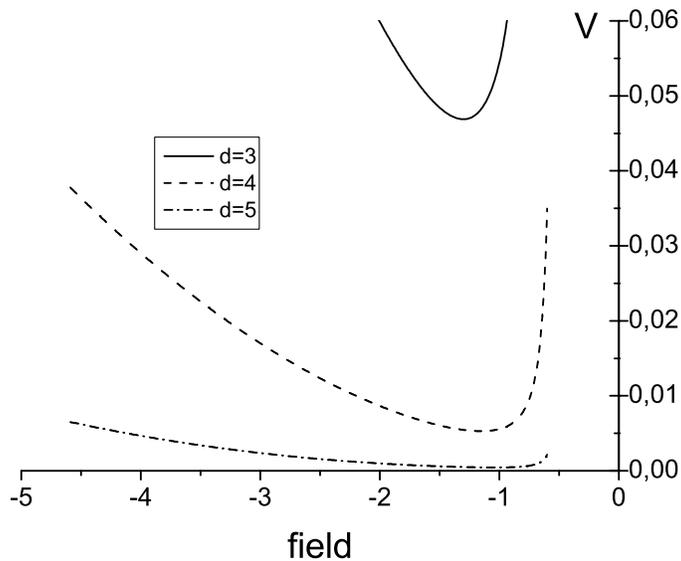}\\
\caption{The effective potential $V(\phi)$ for different numbers of extra
    dimensions for the Lagrangian (\ref{Lgen}) with $F(R)$ of the from
    (\ref{quadr}). Parameter values: $c=1,\ \Lambda =-0.4$.}
    \label{Minimum1}
\end{figure}

\subsection
        {Extra dimensions, inflaton mass and $\Lambda_{\rm eff}$}

    Let us start with discussing the well-known deficiency of chaotic
    inflation in its simplest, quadratic form. According to observations
    of the temperature fluctuations of the cosmic microwave background,
    the inflaton mass is about $10^{-6}\,m_4$. Its smallness needs an
    explanation. To this end, consider the effective potential (\ref{VE})
    generated by the initial action (\ref{Lgen}), (\ref{quadr}). Its shape
    is represented in Fig.\,\ref{Minimum1} for some values of $d$, the
    number of extra dimensions. Evidently, one can fit the parameters in
    such a way that the potential does not contradict the observations. By
    increasing $d$, one could get an arbitrarily small values of the energy
    density at the minimum (at least in the Einstein frame). Its minimum
    can be made shallow to maintain small temperature fluctuation of the
    background radiation.

    Simple numerical calculations give the following numbers: the second-order
    derivative at the bottom of the potential (in fact half mass of the
    inflaton) equals to $\sim 0.2\mD$ for $d=3$; $\sim 1.5\cdot 10^{-3}\mD$
    for $d=5$ and $\sim 0.8\cdot 10^{-5}\mD$ for $d=7$. There are no small
    microscopic parameters in the Lagrangian. Nevertheless, a small
    inflaton mass arises at the classical level, and we can obtain
    appropriate conditions for the chaotic inflation in the early universe by
    choosing a sufficient number $d$ of extra dimensions.
    An estimate of this number depends on the uncertain ratio
    $\mD/m_4$; if $\mD \lesssim m_4$, the number $d = 7$ suits.

    A strong influence of the number $d$ on the effective Lagrangian
    parameters is not surprising since the potential (\ref{VE}) contains a
    quickly decreasing factor $\sim d^{-d}$. Thus, if in a stationary state
    $\phi = \phi_0$, the dimensionless initial parameters $|\phi_0|\mD^{-2}$
    and $F'(\phi_0)$ are of order unity, the effective cosmological constant
    $\Lambdaef = V_{\rm Ein}(\phi_0)$ is related to $F(\phi_0)$ (which may
    be close to $\mD^2$) by
\beq
    \Lambdaef/F(\phi_0) \sim [d(d-1)]^{-d/2}.
\eeq
    It is of interest that $d^{-d} \approx 10^{-123}$ for $d = 67$.
    Thus, at least in the Einstein picture, a fluctuation leading to a
    (67+4)-dimensional space, may evolve to a space with the vacuum
    energy density $10^{-123}\,m_4$. The extreme smallness of $\Lambdaef$ is
    related to the number of extra dimensions, and other physical ideas are
    not required. We have again taken for certainty $\mD = m_4$, otherwise the estimates
    will be slightly different.

\begin{figure}
\includegraphics[width=0.75\textwidth]{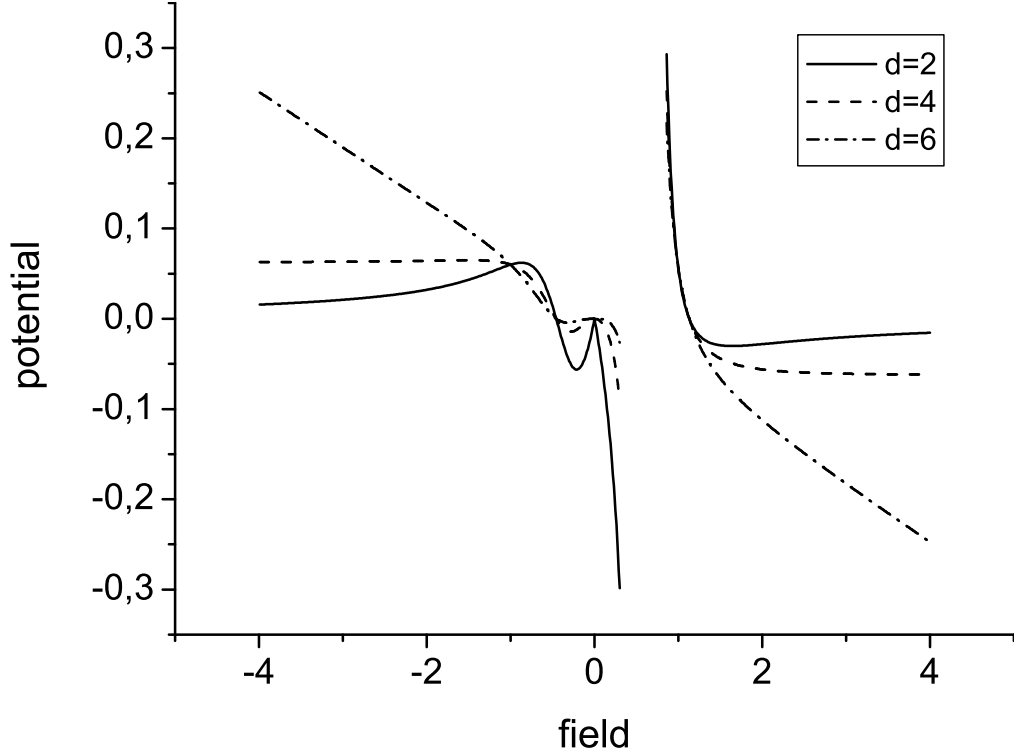}\\
\caption{Different potentials for different $d$, obtained from the
    Lagrangian (\ref{Lgen}) with $F(R)$ in the form (\ref{quartic}).
    Parameters: $c=0,\ w_1 =0,\ w_2 =-1,\ \Lambda =-0.25$.
    The curves are adapted to a unique scale.} \label{d=2-6a}
\end{figure}

    Fig.\,\ref{d=2-6a} gives another example of $d$-dependence of the shape
    of the potential. Even its minimum does or does not exist depending on
    the value of $d$, see the curves at $\phi > 0$.  If a universe is
    nucleated with extra dimensions having negative curvature, we have $\phi
    < 0$, see Fig.\,\ref{d=2-6a}. Evidently the mean value of the potential
    $V(\phi )$ in such a universe tends to infinity for $d=2$, to a
    constant value for $d=4$ and to zero for $d=6$.  All this takes place if
    the initial field value is less than $-1$. Otherwise, if a universe is
    born with $-1<\phi < 0$, it is captured in a local minimum, and the size
    of extra dimensions remains small.

\begin{figure}
\includegraphics[width=0.5\textwidth]{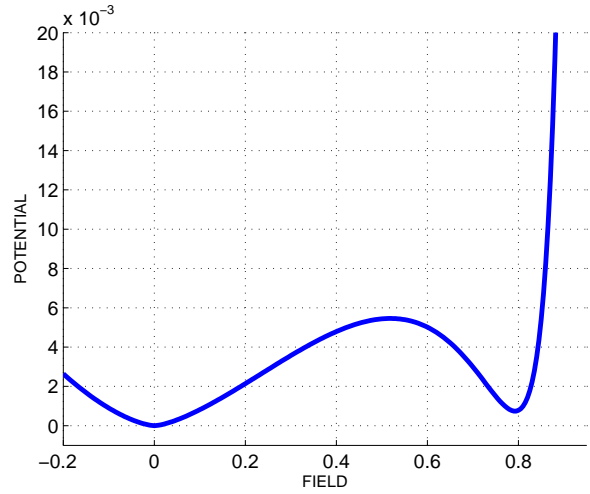}\\
\caption{The effective potential (\ref{VEKr}). The parameters are:
       $c=-0.5,\ \Lambda =0.2,\ c'=-0.626$, and
       $d=2$.} \label{onefield} \end{figure}

    One more example: extra dimensions with the topology of a 3-sphere. The
    effective potential (\ref{VEKr}) for this case is presented in
    Fig.\,\ref{onefield} for positive field values. It has a metastable
    minimum, and a universe residing at this minimum as a ground state could
    live very long. As it was discussed in Section IIIA, such a universe
    could contain primordial massive black holes provided it was formed near
    the top of the potential.

\subsection{Multiple factor spaces and spatially
            dependent size of the extra dimensions.}

    As was mentioned in the Introduction, we do not assume a specific number
    of extra dimensions or their topology. Both are thought to arise due to
    quantum fluctuations close to the Planck scale. If quantum fluctuations
    lead to a more complex structure of the extra dimensions, the physics
    becomes much richer.

    Consider an extra space with two factor spaces: $\M_{D} = \M_{d_{1}}
    \times \M_{d_{2}}$. Treating the same action (\ref{SKr}), we should
    introduce two scalar fields to describe the low-energy limit. The
    effective potential has the following form in the Einstein frame:
\bearr
    V_{\rm Ein}(\phi_1 ,\phi_2 )
    = -\frac12 \sign \left(F'(d_1 \phi_1 + d_2\phi_2 )\right)
        \frac{|\phi_1|^{d_1/2}}{ [(d_1 -1)]^{d_1 /2}}
            \frac{|\phi_2|^{d_2/2}}{ [(d_2 -1)]^{d_2 /2}}
\nnn \inch
    \times \frac{F(d_1 \phi_1 + d_2\phi_2 )
        + d_1 \phi_1^2 [c_1 + 2c_2/(d_1 -1)]
        + d_2 \phi_2^2 [c_1 + 2c_2/(d_2 -1)]}
            {[F'(d_1 \phi_1 + d_2\phi_2 )]^2},  \label{VE2fields}
\ear

\begin{figure}
\includegraphics[width=0.5\textwidth]{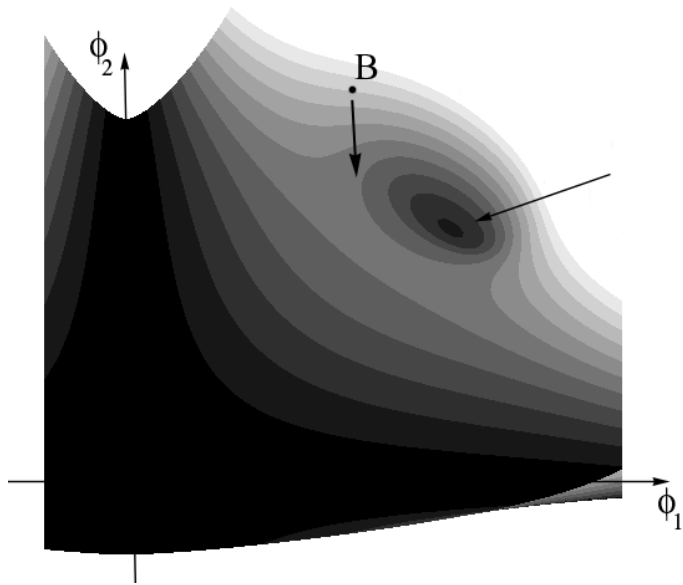}\\
\caption{Effective potential for an extra space with the topology
    $\M_{D}=\M_{d_{1}}\times \M_{d_{2}}$, $d_1 = d_2 = 3$ and the parameters
    are $c=-0.5,\ \Lambda =0.2,\ c_{1}=c_2 =-0.38$.  The view is from above,
    lower levels are darker. The local minimum is marked by a long arrow.}
    \label{TwoFields}
\end{figure}

   It is represented in Fig.\,\ref{TwoFields} where two valleys of the
   potential lie in perpendicular directions, $\phi_1 =0$ and $\phi_2 =0 $,
   each of them corresponding to an infinite size of one of the extra
   factor spaces, $\M_{d_1}$ or $\M_{d_2}$. Of greater interest is the local
   minimum, marked by a long arrow, where both factor spaces are compact and
   have a finite size. A universe can live long enough in this metastable
   state, as in the case of a simpler topology of extra dimensions
   discussed above.

   An interesting possibility arises if a universe is formed at point B in
   Fig.\,\ref{TwoFields}. There occurs inflation, and it ends as the fields
   move from point B along the arrow. The fate of different spatial domains
   depends on the field values in these domains. Even if most of them
   evolve to the metastable minimum, some part of the domains overcomes
   the saddle and tend to the first or second valley, with an infinite size
   of one of the factor spaces. In this case, our Universe should contain
   some domains of space with macroscopically large extra dimension. Their
   number and size crucially depend on the initial conditions.

   The laws of physics in such a domain are quite different from ours. So,
   if, say, a star enters into such a domain, since the law of gravity is
   dimension-dependent, the balance of forces inside the star will be
   violated, and it will collapse or decay. More than that, there will be no
   usual balance between nuclear and electromagnetic forces in the stellar
   matter, so that even nuclei (except maybe protons) will decay as well.
   Even hadrons, being composite particles, are likely to lose their
   stability.

   To conclude, we have seen that the same theory described by a specific
   Lagrangian leads to a diversity of low-energy situations, depending on
   the structure of extra dimensions and the initial conditions.

\section{Conclusion}
\label{Conclusion}

   We have discussed the ability of nonlinear multidimensional gravity to
   produce various low-energy theories, some of which could describe
   our Universe. We only assume a certain form of the pure
   gravitational action and a sufficient number of extra dimensions in a
   Kaluza-Klein type framework, but do not fix the number, dimensions and
   curvature signs of extra factor spaces. Artificial inclusion of matter
   fields is not supposed, so that our conclusions are based on purely
   geometric grounds.

   This problem setting creates a number of promising low-energy theories.
   We have discussed both some new effects and those already known,
   the latter being confirmed in the framework of multidimensional gravity.

   First, we confirm the existence of de Sitter vacua studied in our
   paper \cite{BR06} in a wide range of parameters.

   Second, nontrivial forms of the kinetic term of the inflaton field
   arise naturally in this approach. As a result, inflaton oscillations at
   the end of inflation could be very rapid with an appropriate form of the
   kinetic term. This increases the particle production rate after the end
   of inflation. At the same time, a slow motion of the inflaton at the
   beginning of inflation provides a sufficiently long inflationary stage.

   Third, it has been shown that multiple production of closed walls
   and hence massive primordial
   black holes is a probable consequence of modern models of inflation.
   Their abundance strictly depends on the initial conditions and the input
   parameters of the model. This point must be taken into account in
   any model of inflation containing a potential with at least two minima.

   Fourth, in the framework of multidimensional gravity, the effective
   kinetic term can change its sign, and its nodes are points of special
   interest. It has been shown that such a node can be considered as an
   attractor with a quantum/chaotic behaviour of the scalar field in its
   vicinity. The ground state energy of such a system turns out to
   be time-dependent, revealing a chaotic behaviour near the critical point.

   Fifth, quite different effective low-energy models arise if one considers
   different numbers and/or topology of extra dimensions, even if all
   parameters of the initial Lagrangian are fixed. It means that the
   specific values of these parameters could be less important than it is
   usually supposed: even more important are the number, dimensions and
   curvatures of the extra factor spaces. In particular, varying the
   number $d$ of extra dimensions forming a single factor space, it is quite
   easy to obtain the proper value of the inflaton mass. Moreover,
   the problem of the tiny value of the observed cosmological constant may
   be explained (at least in the Einstein picture) by a moderate number of
   extra dimensions: by our estimate it suffices to have $d = 67$, which,
   though also does not look attractive, can hardly be called an
   ``astronomically large parameter''.

   Sixth, if we take into account that the size of extra dimensions may
   depend on the spatial point in the observed space, our Universe may
   contain spatial domains with a macroscopic size of extra dimensions,
   where the whole physics should become effectively multidimensional.
   Matter (be it a spacecraft, a star or a galaxy) getting into such a
   domain would be unable to survive in its usual form.

   Thus pure multidimensional gravity, even without any other ingredients,
   is quite a rich structure, and many problems of modern cosmology may be
   addressed in this framework. A task of interest is to try to construct
   a model able to solve a number of problems (if not all) simultaneously.

\section{Acknowledgement}
S.R. is grateful for the stimulating environment, hospitality and
partial financial support provided by the Department of Physics of
Manhattan College. K.B. acknowledges partial financial support
from DFG Project No. 436RUS113/807/0-1(R) and Russian Basic
Research Foundation Project No. 05-02-17478.

\end{document}